\documentclass[epj]{svjour}
 \usepackage{latexsym} 
\usepackage{graphics}
\def \b {\begin{eqnarray}}
\def \el#1 {\label{#1} \e}
\def \e {\end{eqnarray}}
\def \ds {\displaystyle }
\def \no {\nonumber\\ }
\newcommand{\ex}[1]{\,\mathrm{e}^ {\ds{#1}}}
\newcommand{\ave}[1]{\langle {#1} \rangle}
\renewcommand{\vec}[1]{\mathbf{#1}}
\newcommand{\abs}[1]{\left| {#1} \right|}

\begin{document}

\title{Molecular velocity auto-correlation of simple liquids observed  by  NMR MGSE method}

\author{Janez Stepi\v{s}nik\inst{1} \and Carlos Mattea\inst{3} \and Siegfried Stapf\inst{3}\and Ale\v{s} Mohori\v{c}\inst{1,2}}

\institute{                    
  \inst{1} University of Ljubljana, Faculty of Mathematics and Physics, Jadranska 19, 1000 Ljubljana, Slovenia, EU\\
  \inst{2} Josef Stefan Institute, Jamova 39, 1000 Ljubljana, Slovenia, EU\\
   \inst{3} Ilmenau University of Technology, Ehrenbergstrase 29, Ilmenau, Germany, EU
}
\date{Received: date / Revised version: date}

\abstract{
The velocity auto-correlation spectra of simple liquids obtained by the NMR method of modulated gradient spin echo  show  features in the low  frequency range  up to a few kHz, which can be explained reasonably well  by  a  $t^{-3/2}$ long time tail  decay  only   for non-polar liquid toluene, while the  spectra of polar liquids, such as ethanol, water and glycerol,  are more congruent with the model of diffusion of particles temporarily  trapped in potential wells created by their neighbors.   As the method provides the spectrum averaged over  ensemble of particle trajectories,  the  initial non-exponential  decay   of spin echoes is attributed  to a  spatial heterogeneity of  molecular motion  in a bulk of liquid, reflected in distribution of the echo decays for short  trajectories. While at longer time intervals, and thus with longer trajectories, heterogeneity is averaged out, giving rise to a spectrum  which is explained as a  combination of molecular self-diffusion  and   eddy diffusion within the vortexes of  hydrodynamic fluctuations.  }


\maketitle

\section{Introduction }
The velocity autocorrelation function (VAF) is a key quantity of the molecular translational dynamics  containing information about the underlying processes of molecular interaction in fluids. Phenomena such as thermal and mass diffusion, sound propagation, transverse-wave excitation, having either a single-particle or a collective nature, are all reflected through the motions of individual particles in the VAF. One of the most significant discoveries  in the field of molecular dynamic in fluids  is the existence of a non-exponential long-time tail (LTT) in VAF, described  for 3D systems by the power law $\approx t^{-3/2}$, at first predicted on the  ground of Landau-Lifshitz  theory of hydrodynamic fluctuation~\cite{Landau,Vladimirsky,Giterman,LandauB}. However, this discovery gained momentum after its confirmation   by   simulations of  hard-sphere fluid  dynamics~\cite{Alder1}, which show  that a diffusing hard sphere  develops   a vortex flow. This flow  is essentially a hydrodynamic back-flow effect responsible for  the persistence of the VAF  at long times \cite{Alder2}. Since then  much experimental,  theoretical,  and computational work has been undertaken in order to understand aspects of this behavior
of the VAF in fluids. 

The  VAF of liquids  can be measured indirectly with neutron ~\cite{Sakamoto,Larsson,Ardente} and light scattering~\cite{Maret}. The short time scale limitation of these  methods  leads to the results that are  not very conclusive on the asymptotic LTT behavior of the VAF~\cite{BoonC}. Some experimental evidence of LTT  is  found in the systems of moderately concentrated polystyrene spheres  and colloidal fluids of hard spheres  by dynamic light scattering~\cite{Boon,Paul,Megen} and by optical microscopy~\cite{Kegel} but without unambiguous determination of its decay. This led to the conclusion that the computer molecular dynamics simulation is the most direct analytical tool for the study of the LTT. 

Theoretical studies  ~\cite{Dorfman} and simulation  for  various systems and molecular interactions ~\cite{Levesque,Levesque2,BoonC,Marro,Rahman,Andriesse,Carneiro,Morkel} reinforced the hypothesis of the power law time dependence of the LTT, but  with  limitations posed by  a finite time interval of simulations and the uncertainty of extrapolation to an infinite number of interacting particles~\cite{Peng}. These studies give comprehensive understanding  of the LTT  in a hard disk/sphere system, in contrast to the systems with more realistic continuous interaction  like a Lennard-Jones potentials, where the LTT appears only  in intermediate densities, almost  in the gaseous state, while in dense systems other dynamical effects on shorter time scales, such as backscattering due to  bouncing of atoms of near neighbours, effectively hide the LTT~\cite{Ryitsev,McDonough,Dib,Bellissima}.

Thus, determination of the VAF asymptotic behavior in dense systems remains a challenge. which complete understanding cannot be  revealed by using  traditional experimental techniques.  Instead, new information  could be provided with  techniques unrelated to these mainstream research tools. A lot of effort has been devoted to understand the molecular translation dynamics in liquids by measuring  the self-diffusion coefficient, $D$ for instances with  tracer techniques and  NMR pulsed gradient spin echo method or by  the direct  measurement of  power spectrum of VAF, i.e the velocity autocorrelation spectrum (VAS) by  modulated gradient spin echo method (MGSE). Especially the latter proved to be very successful at measuring the VAS  of   polymer melts~\cite{moj14_1},  fluidized granular motion \cite{Lasic06}, and  restricted  diffusion in porous systems~\cite{moj001,Codd,Topgaard,Parsons}.

\section{Measurement of molecular  dynamics by NMR gradient spin echo }
 Well known results of  $D$ measurements in water by tracer technique in a wide temperature  range ~\cite{Mills} are commonly used  to calibrate other measuring techniques, particularly  measurements of diffusion by the gradient spin echo method. This method, which  is almost as old as NMR itself~\cite{Hahn,Carr},  uses the magnetic field gradient $\nabla|\bf{B}|=\bf{G}$ (MFG) to detect the translational displacement of molecules via precession of their atomic nuclear spins in magnetic field. The method of pulsed gradient spin echo (PGSE) provides   the spin echo attenuation proportional to the molecular mean squared displacement (MSD)  in the interval between two consecutive  MFG pulses~\cite{Torrey56,Stejskal652}. PGSE measurements of water  $D$ show its follows the Arrhenius law~\cite{Galamba} as well as  the Stokes-Einstein relation  for all temperatures above 273 K, but deviates below it and  in the supercooled regime~\cite{Mallamace}. This was  attributed to the inter-molecular hydrogen bonding~\cite{Lamanna,Chandra}. However, theoretical models  generally predict somewhat larger value of $D$ than experimentally observed~\cite{Mahoney}.   PGSE measurements of water at different  pressures and  temperatures ~\cite{Krynicki,Yoshida} show $D$ with scattered values exceeding the experimental uncertainty~\cite{Mills}. The scattering is commonly assigned to  an  inaccuracy of MFG calibration  or to  convection flows in liquids. However, the measurement of diffusion  in finite time interval of PGSE  may  not result in $D$ according to the  Einstein's definition of $D$, as  the time derivative of MSD)
  in the long time limit~\cite{Einstein},  but in  a time dependent  apparent diffusion coefficient (ADC) according to the Green-Kubo expression~\cite{Kubo,moj14_1}
\b
{D}_{xx}(\tau) = \int_0^{\tau} \ave{
{v}_x(t){v}_x(0)}_{\tau} dt.
\el{gk}
Here ${D}_{xx}$ denotes diffusion along $x$-axis and $ \ave{\dots}_{\tau}$ is   the  ensemble average  over trajectories traveled by particles during the time interval $\tau$.   Both definitions  are  equivalent, if the integral in Eq.\ref{gk} does converge for long $\tau$. Any integral divergence indicates  a time-dependent $D$ containing  information about asymptotic properties of VAF\cite{Visscher}.   ADC may  differ from  $D$ obtained with the tracer method, where an infinite time of observation is assumed. In the  measurements of  liquids, the dependence of ADC  on the interval  between gradient pulses $\tau$, is quite commonly either neglected~\cite{Yoshida} or not observed,  because very short diffusion times are not accessible due to the MFG coil induction. However,  there was a theoretical  attempt to analyze ADC time dependence of  PGSE measurement in liquids  within the frame of hydrodynamic model of Brownian motion~\cite{Lisy}.

 Based on  the general  relation between  the NMR  gradient spin echo attenuation and VAF~\cite{moj81,moj85}, the method of  modulated gradient spin echo  (MGSE) was introduced~\cite{mojcall3,mojcall}. The method  measures directly VAS  at the  frequency determined by the rate of spatial  spin phase modulation. The temporal modulation of the spin phase is created  by the sequence of  radio-frequency (RF) pulses and by pulsing or oscillating MFG.  The  ability of   the techniques with pulsed MFG was demonstrated by  measuring  VAS  of  water flow through porous media~\cite{mojcall3}, and  VAS of the restricted diffusion  in porous media~\cite{moj001,Codd,Topgaard,Parsons}. The technique with  oscillating MFG  shows that   the  resolution  of the diffusion weighted MR images of  brain and the  images of  diffusion tensor of neurons improves with the  increase of the modulation frequency~\cite{Aggarwal}.   However,  the self-inductance of gradient coils limits the  upper frequency range of the technique to below  $1$ kHz. Later on, the MGSE technique with constant MFG was developed, in which the gradient coil self-induction is no longer a limiting factor. The frequency range of the technique is increased  and determined by  the rate of RF-pulses and the magnitude of the fixed MFG. Thus,   the measuring of VAS in the range  above $10$ kHz is possible.  The advantage of the new MGSE technique has been demonstrated by the measurements of the VAS  of  restricted diffusion in  pores smaller than 0.1 $\mu$m \cite{moj16},  the VAS of the granular dynamics in fluidized granular systems \cite{Lasic06} and  by the  discovery of a new low frequency mode of tube motion in melted polymers~\cite{moj14_1}. The method can also employ the  internal  MFG in porous systems, generated by the  susceptibility differences on interfaces,  to obtain information about  the pore morphology  and the distribution of internal MFG ~\cite{moj16}. 
 
\subsection{Modulated gradient spin echo method }\label{MGSEVA}
The MGSE sequence is  basically   a Carr-Purcell-Meiboom-Gill sequence  (CPMG) consisting of initial $\pi/2$-RF-pulse and the train of $N$  $\pi$-RF pulses separated by  time intervals $T$~\cite{Carr,Meiboom}, applied in the background of fixed MFG. CPMG sequence was initially introduced  to reduce the effect of diffusion on measurement of $T_2$ relaxation by shortening the pulse spacing, $T$,   but  the presence of MFG imprints also information on the spectrum of VAF~\cite{moj81,moj85}.  The application of this method for VAS measurements in liquids, requires that consideration be given to other spin interactions besides that with MFG. Although, the rapid molecular motion on the time scale  of pico- or nanoseconds nullifies the spin dipole-dipole and first order quadrupole  interactions  completely,  spin interactions with electrons in the molecular orbitals remain in liquids, resulting in  the chemical shifts of NMR spectrum, and the  electron mediated spin-spin interactions, considered as a J-couplings. Here, we are assuming that  these interactions can be neglected in a strong enough MFG of MGSE sequence. Thus, the Hamiltonian of the spin dynamics  can be simplified to  
\b
{\cal H}&=& -\sum_i{[ \omega_{o}{\cal I}_{zi}+
{\vec\omega}({\vec  r}_i){\cal{\vec I}}_i 
]}+{\cal H}_{rf}(t) + {\cal H}_{L},
\el{ham}  
where  the sum runs over all individual spins. The Zeeman interaction  with the strong external uniform magnetic field oriented along the $z$-axis causes spin precession with the frequency $\omega_o=\gamma B_{0z}$, while the interaction with MFG gives the resonance off-set frequency ${\vec\omega}({\vec  r}_i)=\gamma{\vec G}\cdot{\vec  r}_i$  for the spin at position ${\vec  r}_i$.  RF-term is described according to its effect on spins in the field $B_{0z}$ as ${\cal H}_{rf}(t)={\cal H}_{\pi/2}^x(t)+{\cal H}_{cpmg}^y(t)$. The first term describes the initial  excitation with  $\pi/2$ RF-pulse, which turns the magnetization into the transverse direction along the $y$-axis. The second term  describes the interaction with the train of $\pi$-RF pulses following initial excitation after the interval $T/2$:  ${\cal H}_{cpmg}(t)=-2\omega_{\pi}(t)\cos{( \omega_{o}t)}\sum_i{\cal I}_{yi}$, where $2\omega_{\pi}(t)/\gamma$ denotes the amplitude of  $\pi$-RF pulses. Each $\pi$-RF pulse rotates the magnetization around the $y$-axis for $180$ degrees. The last part of the Hamiltonian, $ {\cal H}_{L}$ includes all other molecular interactions, including the magnetization relaxation~\cite{Kowalewski}.
\begin{figure}
\centering \scalebox{1.3}{\includegraphics{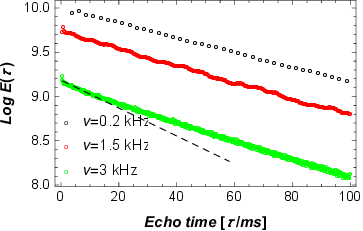}}
\caption{Spin echo decays of  water  at three modulation frequencies, presented  by $N=40, 300$ and $600$ echo peaks respectively,  show a clear deviation from a mono-exponential decay at $\nu=3$kHz .\label{fig1}}
\end{figure}

Complex spin dynamics under the influence of  the sequence of RF pulses and  various magnetic fields  can be solved by  using   Feynman's operator calculus ~\cite{Feynman,Dyson}, in which the Hamiltonian is transformed  into  a series  interaction representations. Subsequent transformations of system,  into the frame of molecular motion with the disentanglement of ${\cal H}_{L}$ in Eq.\ref{ham},  then into the frame rotating with $\omega_{o}$ and finally into the toggling frame determined   by  ${\cal H}_{cpmg}^y(t)$ ~\cite{moj81,mojcall},  amounts into the effective gradient/RF Hamiltonian~\cite{moj16}, which  describes  a combined effect of RF and MFG fields as ~\cite{moj16}
\b
{ \cal H}_{gt}(t)=-\sum_i{\omega_z({\vec  r}_i(t))[{\cal I}_{zi}\cos{b(t)}-{\cal I}_{xi}\sin{b(t)}]}.
\el{efgrad}
Here, $\omega_z({\vec  r}_i)$ is the $z$-component of ${\bf\omega}({\vec  r}_i)$, which includes the   molecular motion, and   the  term with ${\cal I}_{xi}$, which describes the resonance off-set due to simultaneous application of MFG and RF-pulses.  The first term of Eq.\ref{efgrad} changes sign after each $\pi$-RF pulse, because $ b(t)=\int_0^t{\omega_{\pi}(t')dt'}$  toggles  $\cos{b(t)}$ between $\pm 1$.  In the case of  the infinitely short $\pi$-RF pulses, the fast transition between $\pm$-states allows us to  neglect  the second term, but not so if  $\pi$-RF pulses have  finite width  $\delta$. Then   a pulsed perturbation  along ${\cal I}_{x}$  appears during the  transitions. It can affect the MGSE measurement of molecular self-diffusion as shown in the Appendix.

\begin{figure*}
\centering \scalebox{1.1}{\includegraphics{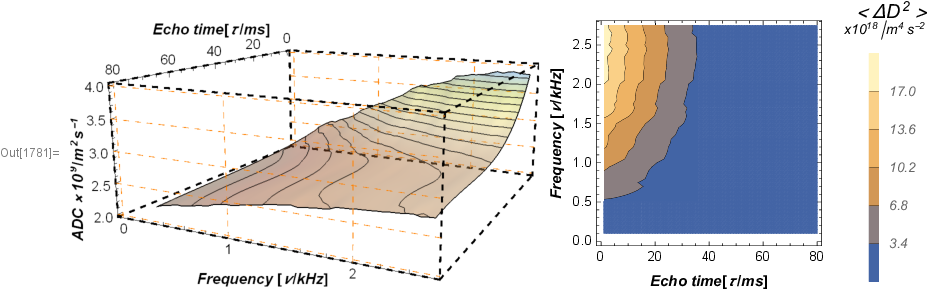}}
\caption{Frequency/temporal plots of ADC  of water (left) and  the variance $<D^2>$ (right) at 23$^{\circ}$C.  The variance levels to zero at long $\tau$ showing that the distribution of  motional properties disappears, when the molecule trajectories are long enough to span  the whole space of heterogeneity.\label{fig2}}
\end{figure*}
\begin{figure*} 
\centering  \scalebox{1.1}{\includegraphics{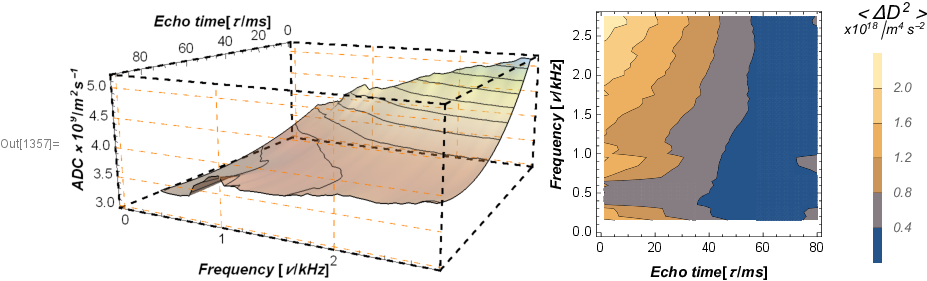}}
\caption{Frequency/temporal variation of   ADC of toluene (left),  and  of  variance $<D^2>$ (right)  at 23 $^{\circ}$C.\label{fig3}}
\end{figure*}

In the case of  $\pi$-RF pulses  short enough that $ \Delta\phi<1$ (see Appendix) and negligible spin relaxation, the first approximation of  the spin echo amplitude induced by the $y$-component of magnetization at echo time $\tau=2 N T$ is  
\b
 E(\tau) &\approx&{d\over dt}\sum_i Tr\, \rho (\tau) {\cal  I}_{yi}\no&\approx&\sum_i{\ave{ \ex{i \int_0^{\tau}\omega_z({\vec  r}_i(t))\cos{b(t)}dt}}},
\e
if  uniform sensitivity of the receiver coil across the excited volume of sample is assumed~\cite{Ales09}. Here, the oscillation  of  $f(t)=\int_0^t\cos{(b(t')}dt'$ permits the integration par parts to write 
\b
E(\tau)&\approx&\sum_i{\ave{ \ex{-i\int_0^{\tau}\nabla\omega_z({\vec  r}_i(t)\cdot{\vec v}_i(t)f(t)dt}}},
\el{sig}
where $\vec{v}_i(t)$ is the velocity of the tagged spin bearing  particle at time $t$. 

Generally,   fluctuations of a molecular system in the thermal equilibrium  are characterized by  correlation functions of relevant physical quantities. Here, we assume that MFG is strong enough  that  only fluctuation of molecular translation velocity,  $\Delta {\vec  v}_i(t)= {\vec v}_i(t)- \ave{ {\vec v}_i(t)}$,  can be  taken into account. With the assumption that  $\Delta {\vec  v}_i(t)$ is a random variable,   Eq.\ref{sig} can be expanded  into the cumulant series.  In case when the spatial discord of spin phase created by  MFG,  i.e. $\lambda=2\pi/\gamma GT$, is larger than the spin displacements in the interval $T$, the first   two   terms  of expansion  are sufficient for the Gaussian phase approximation, which gives the spin echo amplitude~\cite{moj81,mojcall,moj202} 
 \b
 E(\tau)&=&\sum_iE_{oi}\ex{i\alpha_i(\tau)-\beta_i(\tau)},
\e
where the sum goes over the sub-ensemble of spins that have  the same dynamical properties. In the case of molecular diffusion in homogeneous media,  the first phase shift term  $$\alpha_i(\tau)=\int_0^{\tau}\nabla\omega_z({\vec  r}_i(t))\cdot\ave{{\vec v}_i(t)}f(t)dt$$ is canceled after a few CPMG cycles due to  toggling effect of  $f(t)$.  However, This is not true in the case of  the restricted diffusion~\cite{moj202}. The second term describes the  echo attenuation, which can be expressed in  the frequency domain as~\cite{mojcall}
 \b
\beta_i(\tau)&=&\frac{1}{\pi}\int_{0}^\infty \vec{q}( \omega ,\tau )\vec {D}_i (\omega ,\tau) \vec{q}^* (\omega ,\tau ) \,d\omega.
\el{att}
  Here the spectrum of the spin phase discord ${\vec q}(\omega,t)=\nabla\omega_z({\vec r}_i)f(\omega,t)$ is determined by  $ f(\omega,t)$, which is the frequency spectrum of $f(t)$~\cite{moj81,moj85}.  $\vec {D}_i $ is the tensor of the VAS 
\b
\vec{D}_i(\omega,\tau) =\frac{1 }{\pi} \int_{-\infty}^{\infty} \ave{\Delta
\vec{v}_i(t)\otimes\Delta\vec{v}_i(0)}_\tau e^{\displaystyle{-i\omega t}} dt,
\el{psd}
  After, a few CPMG cycles, $(N>4)$~\cite{mojcall3}, the power  spectrum  of $f(t)$ can be  approximated  by
\b
\abs{f(\omega,t)}^2&\approx&2\pi t\sum_{k=-\infty}^{\infty}{\abs{c_k}^2\delta(\omega-k\omega_m)},
\el{fspec} 
in which narrow lobes with the amplitude \\$\abs{c_k}^2= 4\sin{(k\pi/2)}^2/k^4\pi^2\omega_m^2$ appear  at the  multiples of the modulation frequency, $\omega_m=\pi/T$ . Neglecting  all   but the dominant first term and with  MFG  applied along the $z$-axis, the  echo peak amplitude at $\tau=NT$ can be written as
\b
E(\tau,\omega_m)&=&\sum_i{E_{oi}\ex{-{\tau\over T_{2i}}-\frac{8\gamma^2G^2}{\pi^2 \omega_m^2}D_{zzi}(\omega,\tau)\tau}}, 
\el{dusapprox}
if spin relaxation is included. Here  $D_{zzi}(\omega,\tau)$ is the $z$-projection of the VAS tensor  averaged  over the trajectories of molecular motion in the interval $\tau$.  The described method provides the low frequency part of the VAS  by changing the modulation frequency  in the range, which is  limited above  by the power of RF transmitter.
\begin{figure}
\centering \scalebox{0.9}{\includegraphics{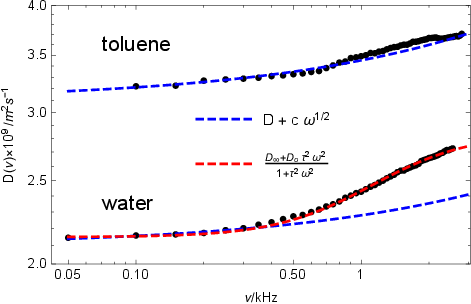}}
\caption{The VAS of water and toluene at 23 $^{\circ}$C obtained at $\tau>50$ ms, when   the spin echo decays  mono-exponentially. Red dashed lines show the fit by Eq.\ref{coup}, while the blue dashed  lines show the fit by the anticipated $\sqrt{\omega}$-dependence of spectrum corresponding  to  $t^{-3/2}$-LTT dependence of VAF. It  matches reasonably well only for VAS of toluene.  \label{fig4}}
\end{figure}

\section{Experiments}
Measurements  were carried out on two different systems: TecMag $100$  MHz NMR spectrometer with a $2.35$ T  superconducting magnet equipped with the Maxwell gradient coils capable of generating MFG in steps to $5.7$ T/m. Its high field  allows precise MGSE  measurement of VAS in liquids, but low MFG  limits the frequency range to the interval from $50$ Hz to $3000$ Hz. Spin relaxation contribution  is determined by separate measurements in zero MFG. Much higher but constant MFG of  the  NMR-MOUSE~\cite{Blumich} of $21.6$ T/m allows measurements in the frequency interval from $50$ Hz to $10$ kHz even  of such slow self-diffusion  as that of  glycerol. 

In order to test the effect of molecular  interaction between  nearest neighbors in dense fluid that could effectively hide the LTT~\cite{Ryitsev,McDonough,Dib,Bellissima}, we conducted MGSE measurement  on three polar liquids: distillate  water, ethanol (analytical standard-Sigma-Aldrich), and  glycerol (99.5$\%$-Sigma-Aldrich)  with the dipole moments at room temperature of $1.85$ D, $1.69$ D, and $2.56$ D respectively and non-polar  toluene (99.8 $\%$-Sigma-Aldrich). The interference of restriction to diffusion by sample boundaries was eliminated by enclosing the  samples in a cylindrical glass cell, $15$ mm long and with $5$ mm  diameter, which are much larger than  molecular  displacements in the interval of measurement.

 The results were  checked  with the measurements of all samples on both systems, but with a lower  accuracy  on the   NMR-MOUSE due its low proton Larmor frequency of $18.7$ MHz.

\section{Results and discussion}

\begin{figure*}
\centering \scalebox{1.2}{\includegraphics{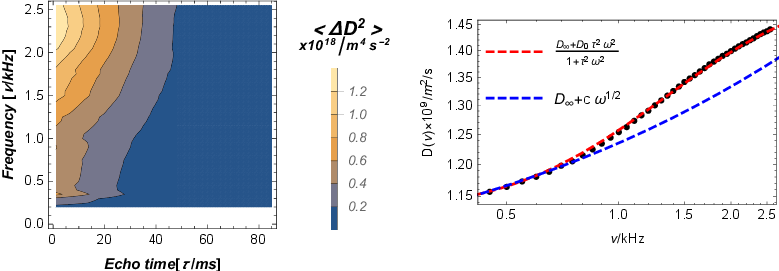}}
\caption{Frequency/temporal plots of  the variance $<D^2>$ (left)  and VAS  of ethanol (right) at 23 $^{\circ}$C obtained from the  spin echo attenuation  in the range of mono-exponential decay. Red dashed line shows a good  fit by Eq.\ref{coup}, but only a few points at the low frequencies can be  approximately fitted  with LTT-dependence (the blue dashed). \label{fig5}}
\end{figure*}

Preliminary MGSE  measurements  of various  liquids (water, ethanol, toluene, hexane, glycerol and water /glycerol mixtures) reveal unexpected  low frequency features  of VAS.  Here  we will only  focus on the results of water, toluene, ethanol and glycerol. In each experiment a train of echoes was recorded with increasing $\tau$ (also $N$) and normalized for transverse relaxation. Experiments were repeated with different $T$ (thus changing the modulation frequency $\nu=1/2T$). Log of spin amplitude vs decay time $\tau$ is a line with the slope  proportional to $D(\omega_m)$ according to Eq.\ref{dusapprox}. However, at short $\tau$ a nonlinear decay of attenuation is observed as shown in Fig.\ref{fig1}. In a heterogeneous system, like liquid in porous medium, this is commonly attributed to the distribution of decay times. By grouping the spins into separate sub-ensembles corresponding to  spins  in  different regimes of diffusion or in a different internal MFG~\cite{moj16} one is able to distinguish groups of spins, which have differing starting points for their motion. In the case of short decay times $\tau$, the  particle displacements  are so small that their trajectories do not sample the entire space, and  can convey information about   heterogeneity of translational  dynamics~\cite{Swallen}. The contribution of sub-ensembles  with the distribution $P(D)$ to the recorded signal  $E(\tau)=\int P(D) e^{-s D\tau} dD$  gives rise to non-exponential decay. With a small deviation from linearity the spin echo attenuation can be approximated by
\b
\beta(\tau)=\log[E(\tau)]\approx -\tau/T_2-s\ave{D}\tau+\frac{s^2}{2}\ave{\Delta D^2}\tau^2+...,
\e
 where $\ave{ D}$ is the mean  diffusion coefficient, $\ave{\Delta D^2}$ is the variance of distribution and $s= \frac{8\gamma^2G^2}{\pi^2\omega^2}$ is  given in Eq.\ref{dusapprox}. Fitting the  spin echo decays  obtained by the measurements in water, toluene, and ethanol by the polynomial  of the fifth order gives the curves  with  the coefficient of determination $R^2>0.999999$. After normalization for the spin relaxation time,  first derivatives of  $\beta(\tau)$ give the values, which we considered as ADCs. Their dependence  on the spin-echo time $\tau$ and  on the frequency of modulation $\nu=1/2T$ are presented  in  a frequency/temporal  3D plots  in Fig.\ref{fig2} for water, and   in Fig.\ref{fig3} for toluene.  Both figures are supplemented by the frequency/temporal  contour plot of  second derivatives of $\beta(\tau)$, which describe  the variance of diffusion distribution, $\ave{\Delta D^2}$.  Figures show that at  short $\tau$ and at high modulation  frequencies there is a non-zero value of variance, which  reflects local diversity observed when particle trajectories are short enough.  Figures also show that at $\tau$ longer than $40$ ms  and   frequencies below $50$ Hz, when particle trajectories are long enough to span whole space of heterogeneity,  $\ave{\Delta D^2}$ levels to zero  and  ADC becomes  equal to  VAS of the liquid. The same is shown for the ethanol in Fig.\ref{fig5}.  From the frequency dependence of $\ave{\Delta D^2}$, we can estimate the size of heterogeneity to about a few tens of micrometers, which agrees with  predictions~\cite{Mazur}.

 When molecules are packed together in liquids, the attractive/repulsive interactions between neighbors exert  effects that could be reflected in the VAS of liquids as  was already confirmed by the  simulation studies~\cite{Ryitsev,McDonough,Dib}. In the case of  polar liquids like water, ethanol, and glycerol  impact of this interactions on   the VAS  should be much greater than for a  non-polar toluene  The results of measurement in toluene, shown in Fig.\ref{fig3},  exhibit a non-exponential decay at short $\tau$, but data obtained  from the  interval of   mono-exponential decay ( $\tau> 40$ ms) give the VAS of toluene, changing from $3.60\times10^{-9}$ m$^2$/s at $40$ Hz to  $4.45\times10^{-9}$ m$^2$/s at $3$ kHz.  The anticipated $\sqrt{\omega}$-dependence, corresponding  to  $t^{-3/2}$-LTT of VAF,  can be roughly fitted  to the obtained the VAS of toluene with the deviation  within the experimental error, as  shown in Fig.\ref{fig4}.  While the VAS of water, obtained in the interval of mono-exponential decay, which  increases  from $2.2\times10^{-9}$ m$^2$/s at $50$ Hz to  $2.75\times10^{-9}$ m$^2$/s at $3$ kHz, and the same of ethanol, which   increases  from $1.15\times10^{-9}$ m$^2$/s at $40$ Hz to  $1.45\times10^{-9}$ m$^2$/s at $3$ kHz,  show strong deviation from   $\sqrt{\omega}$ as shown on Figs.\ref{fig4} and \ref{fig5}. These figures show  attempts  to fit  initial few points  in the frequency range below $500$ Hz for water  and  below $900$ Hz for ethanol  with $\sqrt{\omega}$-curve, which show a strong  deviation from experimental data at higher frequencies..

Self-diffusion coefficients in these liquids mostly  follow the Arrhenius law~\cite{DouglassVFC}, which means that the  molecules in  liquid state are momentarily trapped in potential wells created by their neighbors. These measurements gave  an averaged  excitation potential for water of $18$ kJ/mol while for non-polar toluene of  about $8$ kJ/mol. We assume that  molecular motions and hydrodynamic fluctuations reduce the potential well enough to allow approximation of the molecular dynamics with the set of  Langevin equations of particles harmonically  coupled in pairs. Its solution gives  the low frequency part of VAS in the form
\b
 D_c(\omega)=\frac{D_{\infty}+D_o\tau_c^2\omega^ 2}{1+\tau_c^2\omega^ 2},
 \el{coup}
if  the  inertial terms  are neglected. Here,  $D_{\infty}$ is an averaged diffusion rate, which depends on the number of coupled molecules,  and  $D_o$ is the diffusion rate of  a molecule escaping the binding.  $\tau_c$ is the correlation time, which shortens with the strength of binding. Fig.\ref{fig4} and Fig.\ref{fig5} show   fits of  $D_c$ from Eq.\ref{coup}  to the obtained data  for  VAS of water and ethanol with the coefficient of determination of $R^2>0.99994$.  

\begin{figure}
\centering \scalebox{1.1}{\includegraphics{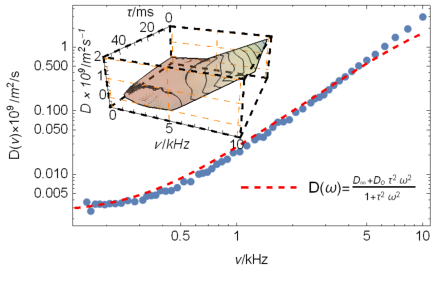}}
\caption{VAS of  glycerol at 23 $^{\circ}$C obtained by NMR MOUSE  fits well with the formula \ref{coup} (red dashed line), while 3D inset  shows the absence of  temporal variation.  \label{fig6}}
\end{figure}

 NMR MOUSE, device with a strong constant MFG and low Larmor frequency $\omega_o$, is not suitable for the measurements of water, ethanol and toluene below $1$ kHz, because of excessive attenuation at low frequencies, i.e. at long $T$. Nevertheless, their VAS  at higher frequencies match well those from $100$  MHz NMR spectrometer.  But the strong MFG of NMR MOUSE suits well for the measurement of very low diffusion coefficient of  glycerol, a substance with large dipole moment and strong hydrogen bondings. The inset  in  Fig.\ref{fig6} shows  that ADC of glycerol does not exhibit any dependence on the echo times. This means that unlike  water and alcohol, there is no effect of  distribution of diffusion on evolution times. It allows a direct  exponential fitting of its echo decays providing  the VAS of glycerol with the coefficient of determination of $R^2\approx 0.9996$ as shown in Fig.\ref{fig6}. It changes  from the  initial value of  $3\times10^{-13}$ m$^2$/s at $50$ Hz to  $10^{-9}$ m$^2$/s  at $10$ kHz. The data can be fitted by Eq.\ref{coup}, but with a lower coefficient of determination as shown in Tab.\ref{tab1}. 

 The asymptotic values of  VAS shown in Figs.\ref{fig4}, \ref{fig5} and \ref{fig6} to the zero frequency,  which are equal to the diffusion coefficient in accordance with the Einstein's definition in Eq.\ref{gk}, for all liquids, correspond  to the values obtained from measurements on other devices~\cite{Mills,Pickup,Meckl,DErrico}.

\begin{table}
\caption{Parameters of Eq.\ref{coup} fitting to VAS }
\label{tab1} 
\begin{tabular}{lllll}
\hline\noalign{\smallskip}
sample & $D_{\infty}$& $D_o/10^{-9}$ m$^2$s$^{-1}$& $\tau_c$/ms&$R^2$  \\
\noalign{\smallskip}\hline\noalign{\smallskip}
Ethanol &$1.12$ & $ 1.54$ & $0.11$&$0.99996$\\
Water & $ 2.19$ & $2.9$ & $0.12$&$0.99996$\\
Glycerol & $0.0025$ & $5$  & $0.01$&$0.940$\\
\noalign{\smallskip}\hline
\end{tabular}
\end{table}
\section{Conclusion}
In  conclusion we can state that  MGSE measurements  unveil   the low frequency  VAS of simple liquids, which   approximately endorse the LTT origin predicted by the theory and simulations only for  toluene, while in the cases of  ethanol, water and glycerol, low-frequency VAS can be better explained by    self-diffusion of molecules  temporarily  trapped in potential wells created by their neighbors in their course of motion. Even in the case of toluene, the deviation from the LTT dependence that is within the experimental error could be attributed to interactions, which are small compared to polar fluids, which also is proven by   the smallness of its diffusive excitation potential of $8$ kJ/mol.

Spin-echo non-exponential  decay   in the initial short time intervals, unambiguously confirms an existence of  diversity of diffusion in bulk of liquids, which is not for instance the consequence of media in-homogeneity as  observed in the diffusion measurement in the porous system. Given that the method provides the ensemble average of VAS over the particle trajectories in  the time interval of measurement,  the transition  into mono-exponential spin-echo decay at longer times means that   long enough trajectories, which  traverse entire space of in-homogeneity,  conveys  VAS averaged  over the space of heterogeneity.

The inset in  Fig.\ref{fig6} shows that  ADC of  glycerol does not exhibit any dependence on echo times. We  speculate that the observed the variance of diffusion distribution in the bulk of liquids is associated with hydrodynamic fluctuations, in which the molecular diffusion process is affected by  the  vortex motion of fluid~\cite{Alder2}. Thus, in addition to the size of heterogeneity, the rate of its change  is also important.  Thus, we can explain the absence of  $\ave{\Delta D^2}$  in the glycerol   by the fluctuation of hydrodynamic  vortexes, with rate  too fast to be observed  in the shortest intervals $T$ attainable by our devices. According to the derivation of Eq.\ref{coup}, a shorter $\tau_c$ means also  stronger interactions. Tab.\ref{tab1} shows the correlation times, among which $\tau_c$ of glycerol is  ten times shorter than those of water and ethanol, which  corresponds to the magnitude of dipole moment of these liquids.  

 So we can conclude that MGSE measurements provide information of molecular translation dynamics in liquids, which can be explained as  a combination of molecular self-diffusion and   eddy diffusion processes~\cite{Graham} in the vortexes of  hydrodynamic fluctuation,  while LTT is effectively hidden by intermolecular interactions in polar liquids.

\section{Acknowledgement}
We acknowledge the contribution of  prof. dr. Igor Ser\v{s}a and dr. Franci Bajd from Josef Stefan Institute  to assist in the preparation of experiments and Slovenian research agency, ARRS, for the financial support in the program P1-0060.

\section{Appendix}
Given that ${ \cal H}_{gt}(t)$ is periodic  and cyclic  with period $2T$, the resonance off-set effect can be calculated by using the  Magnus expansion of the time evolution operator~\cite{Magnus}.  The expansion  to the   third order of the cumulant series gives  the averaged Hamiltonian,  whose  first term has been already considered  in Sec.\ref{MGSEVA}.  The derivation of  the second  and  the third terms  results in the Hamiltonian 
\b
{ \cal H}_{gtc}=-(1-\delta/2T)\frac{\delta}{\pi}\sum_i{\omega({\vec  r}_i)^2[{\cal I}_{yi}-\frac{\delta}{\pi}\omega({\vec  r}_i){\cal I}_{zi}]},
\e
which describes   the resonance off-set distortion over a cycle $2T$  as an additional spin rotation in the interval of $\pi$-RF pulse application.  Straightforward calculation results in a factor effecting the signal  induced by the $y$-component of the $i$-th spin sub-ensemble as
\b
k_{ic}=\frac{1+\Delta \phi _i^2 \cos \left(\pi \tau(1-\delta/2T) \Delta \phi _i^2 \sqrt{1+\Delta \phi _i^2}/\delta\right)}{1+\Delta \phi _i^2},
\e
where  $ \Delta\phi_i=\frac{\delta}{\pi}\omega({\vec  r}_i)$. Summation  over the sub-ensembles in the volume, which is  either selected by the  initial  $\pi/2$-RF pulse excitation in the background of   MFG or determined by the size of the sample container~\cite{moj16}, gives  the reduction factor of the echo amplitude  
\b
k_c=\frac{2}{\Delta \phi}\arctan{\Delta \phi/2}
\e
with  $ \Delta\phi=\frac{\delta}{\pi}\gamma\abs{{\vec G}\cdot\Delta{\vec  r}}$, where $ \abs{\Delta{\vec  r}}$ is the  width of  the excited spin slice.  Added to the signal is also a small oscillation with the amplitude   $ {\Delta \phi}^2/12$  and   the frequency  ${\Delta \phi}^2/2\delta$ as shown  in Fig.\ref{fig1}.  Therefore, the  interval of observed echo decay should be much longer than the inverse frequency of oscillations, $\tau>2\delta/{\Delta \phi}^2$ in order to obtain  correct information about  molecular motion from the  average over the  signal oscillations.  The  spin excitation by initial  $\pi/2$-RF pulse  in the background of  constant  MFG  provides the  active volume giving  ${\Delta \phi}\approx 1$, which means that averaging over the time of several $\delta$  is sufficient to suppress resonance off-set distortion. 

At the end of this discussion, it is necessary to address the description of the spin echo by the concept of coherence pathways, which is sometimes incorrectly used in MGSE measurements. Accordingly, the  contributions of different  coherence pathways  to  the shape and decay of signal  differently depend on  diffusion and relaxation~\cite{Hurlimann2,YQSong,Toumelin}.  By the  frequency filtering  of the spin echo signal certain coherence pathways can be excluded so that we get the one that best suits  the  diffusion and relaxation measurements. The zero frequency filtered  echo signal, which isolates the direct coherence pathway, is considered as the best to provide credible information about molecular diffusion. This is true if we assume the self-diffusion coefficient which is constant  in the interval of measurement. In the case of a slow molecular motion, the signal filtering removes  information about motion that is above the threshold frequency of low pass filter as is proven by measurements of water in the reference~\cite{Sersa}.  Thus, the straightforward  calculation of the time integral over the spin echo is by definition the zero frequency component spin echo, in which all information about  the spin motion with  frequencies $\omega_c>\gamma\abs{{\vec G}\cdot\Delta{\vec  r}}/2$ are filtered out.  Thus, it is important to take into account  the principle of  diffusion measurement by the gradient spin echo, where only the peak of the spin echo, which  is by definition the integral over the whole echo  spectrum, $E(N\tau)=\int E(\omega)d\omega$,  conveys information about molecular translational motions~\cite{Stejskal651,moj81,mojcall}, while the Fourier transformation of  the spin echo signal  gives the image of the spin spatial distribution.  This is similar to the uncertainty principle in the quantum mechanics: \lq\lq{}The more precisely the position  of a particle is given, the less precisely can one say what its momentum is.\rq\rq{}. 

\section{Authors contributions}
J. Stepi\v snik, C. Mattea and A. Mohori\v c were involved in the measurements and the preparation of the manuscript, while S. Stapf has read and approved the final manuscript.

\end{document}